\documentclass[prc,aps,superscriptaddress,twocolumn,showpacs,nofootinbib]{revtex4}

\usepackage{graphicx,amsmath,amssymb,bm,multirow}
\usepackage{amsthm}
\usepackage{amsfonts}
\usepackage{dcolumn}
\usepackage{bm}
\usepackage[usenames,dvipsnames]{color}

\begin{document}
%
\title{Nuclear quantum shape-phase transitions in odd-mass systems}

\author{S. Quan}
\affiliation{School of Physical Science and Technology, Southwest University, Chongqing 400715, China}
\author{Z. P. Li}\thanks{zpliphy@swu.edu.cn}
\affiliation{School of Physical Science and Technology, Southwest University, Chongqing 400715, China}
\author{D. Vretenar}
\affiliation{Physics Department, Faculty of Science, University of Zagreb, 10000 Zagreb, Croatia}
\author{J. Meng}
\affiliation{State Key Laboratory of Nuclear Physics and Technology, School of Physics, Peking University, Beijing 100871, China}
\affiliation{Yukawa Institute for Theoretical Physics, Kyoto University, Kyoto 606-8502, Japan}
\affiliation{Department of Physics, University of Stellenbosch, Stellenbosch, South Africa}

\begin{abstract}
Microscopic signatures of nuclear ground-state shape phase transitions in odd-mass Eu isotopes are explored starting from excitation spectra and collective wave functions obtained by diagonalization of a core-quasiparticle coupling Hamiltonian based on energy density functionals. As functions of the physical control parameter -- the number of nucleons -- theoretical low-energy spectra, two-neutron separation energies, charge isotope shifts, spectroscopic quadrupole moments, and $E2$ reduced transition matrix elements accurately reproduce available data, and exhibit more pronounced discontinuities at neutron number $N=90$, compared to the adjacent even-even Sm and Gd isotopes. The enhancement of the first-order quantum phase transition in odd-mass systems can be attributed to a shape polarization effect of the unpaired proton which, at the critical neutron number, starts predominantly coupling to Gd core nuclei that are characterized by larger quadrupole deformation and weaker proton pairing correlations compared to the corresponding Sm isotopes.
\end{abstract}

\maketitle



Quantum mechanical systems can undergo zero-temperature phase transitions upon variation of a non-thermal control parameter. Quantum phase transitions (QPTs) present a very active field of research and have found a variety of applications in many areas of physics and chemistry \cite{Sachev00,Carr10}. Nuclear QPTs, in particular, correspond to shape transitions  between competing ground-state phases induced by variation of a non-thermal control parameter (number of nucleons) \cite{Iac.03,Rick.06,Rick.07,CJ.09,CJ.10}. Most experimental and theoretical studies of first- and second-order nuclear QPTs have considered systems with even numbers of protons and neutrons \cite{FI.00,CZ.00,FI.01,CZ.01,Pietralla04,Jolie01,Meng05,Niksic07,Li09a,Li09b,Li10,Li13}. QPTs in odd-A nuclei present a more complex phenomenon because of the coupling between single-particle and collective degrees of freedom.  The crucial issues for QPTs in odd-A systems are the influence of the unpaired fermion(s) on the precise location and nature of the phase transition, empirical signatures of QPTs, and the definition and computation of order parameters \cite{FI11,Petrellis11}. In recent years phenomenological geometric models with single- or multi-$j$ state coupling \cite{FI.05,Alonso07,Zhang10,Zhang11}, the interacting boson-fermion framework~\cite{Alonso07,Boyu10,Jafa15}, and microscopic energy density functionals~\cite{Nomura16,Nomura17} have been employed in extensive studies of QPTs in odd-mass nuclei. 

In this paper we report a microscopic study of QPT in odd-mass Eu isotopes, calculate a series of observables  that can be related to order parameters  (low-energy spectra, two-neutron separation energies, isotope shifts, spectroscopic quadrupole moments, and reduced transition matrix elements), both for odd-mass nuclei and the adjacent even-even isotopes, and analyze the polarization effect of the unpaired nucleon on the QPT. The choice of Eu isotopes is motivated by the fact that probably the best example of a QPT in atomic nuclei is in the rare earth region with $N\approx90$ neutrons, where a transition between spherical and axially symmetric equilibrium shapes has been extensively investigated both experimentally \cite{CZ.01,Krucken02,Tonev04,Moller06,Kulp08}, and by using a number of theoretical methods \cite{McCu04,Meng05,Niksic07,Li09a,Li09b}. Moreover, the QPT in the odd-proton and even-neutron Eu isotopes is determined by the same control parameter, that is, the number of neutrons, as in the adjacent even-even Sm and Gd isotopes.

Our model is based on the nuclear covariant density functional theory (CDFT)~\cite{Ring96,Vret05,Meng06,Meng16}, specifically the relativistic Hartree-Bogoliubov (RHB) implementation of the CDFT framework, which has successfully been applied to the description of a variety of structure phenomena over the entire chart of nuclides. Modelling excitation spectra and electromagnetic transition rates requires including correlations beyond the static mean-field through the restoration of broken symmetries and configuration mixing of symmetry-breaking product states.  In the present analysis we employ a generalized five-dimensional collective Hamiltonian (5DCH), with quadrupole deformations as dynamical collective coordinates for the even-A system. The microscopic self-consistent solutions of deformation-constrained triaxial RHB calculations: the single-particle wave functions, occupation probabilities, and quasiparticle energies, are used to calculate the Hamiltonian parameters. The resulting collective potential and inertia parameters as functions of collective coordinates determine the dynamics of the 5DCH \cite{Li10b,NVR.11}. For the odd-mass system, we add a quasiparticle to 5DCH and construct a microscopic core-quasiparticle coupling (CQC) Hamiltonian, for which the collective degrees of freedom of the core and the fermion degrees of freedom of the quasiparticle are described within the same CDFT \cite{Quan17}. The inclusion of both neighbouring even-even core nuclei in the CQC Hamiltonian enables the model to take into account shape polarization effects, that is, differences in shapes and related observables between two cores, which are critical for transitional nuclei. The CQC Hamiltonian predicts excitation energies, kinematic and dynamic moments of inertia, and transition rates that are in very good agreement with experiments for deformed odd-mass nuclei \cite{Quan17}.

\begin{figure}[hth!]
\centering
\includegraphics[width=8.2cm]{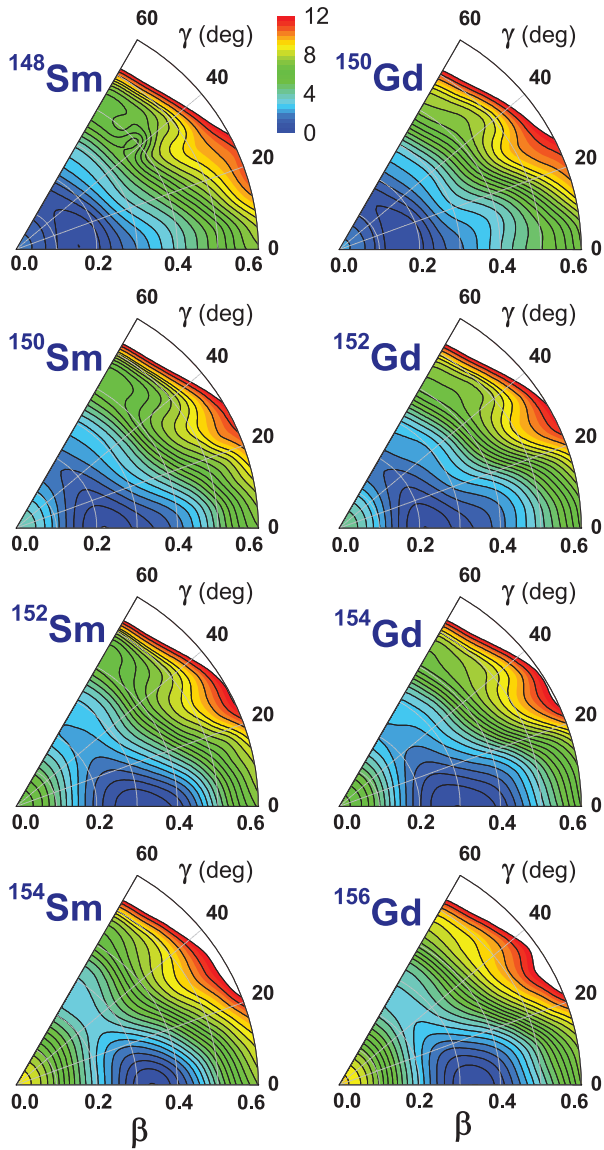}
\caption{\label{fig:PES}(Color online)  Self-consistent RHB triaxial quadrupole energy surfaces in the $\beta$-$\gamma$ plane ($0\le \gamma\le 60^0$) for Sm and Gd isotopes. All energies are normalized with respect to the binding energy of the corresponding ground state. The contours join points on the surface with the same energy, and the separation between neighboring contours is 0.5~MeV.}
\end{figure}

The present analysis starts with the calculation of total energy surfaces as functions of quadrupole deformation coordinates for the even-even Sm and Gd isotopes, using the constrained RHB model based on the PC-PK1 density functional \cite{Zhao10} in the particle-hole channel, and a finite-range separable pairing force~\cite{TMR09} in the particle-particle channel. The deformation energy surfaces are displayed in Fig.~\ref{fig:PES}, and  exhibit a distinct evolution of prolate deformation with increasing neutron number, from the nearly spherical $^{148}$Sm and $^{150}$Gd, to the well-deformed prolate $^{154}$Sm and $^{156}$Gd, as well as the reduction of the $\gamma$-dependence of the potentials. The energy surfaces of $^{152}$Sm and $^{154}$Gd indicate that these are transitional nuclei, characterized by a softer potential around the equilibrium minimum in the $\beta$ direction. Therefore, with increasing $N$ the shape evolution in Sm and Gd isotopes undergoes a QPT between the vibrational and rotational limits of the Casten symmetry triangle \cite{Rick.06}, with $^{152}$Sm and $^{154}$Gd being located closest to the critical point.

Even though shape coexistence and transitions in nuclei have been extensively explored by considering potential energy surfaces, a quantitative study of QPT must go beyond the simple Landau approach and include direct computation of observables related to order parameters. In the following we will consider spectroscopic properties of odd-mass Eu isotopes that can be associated with order parameters of a shape phase transition. 

\begin{figure}[htb!]
\centering
\includegraphics[width=9cm]{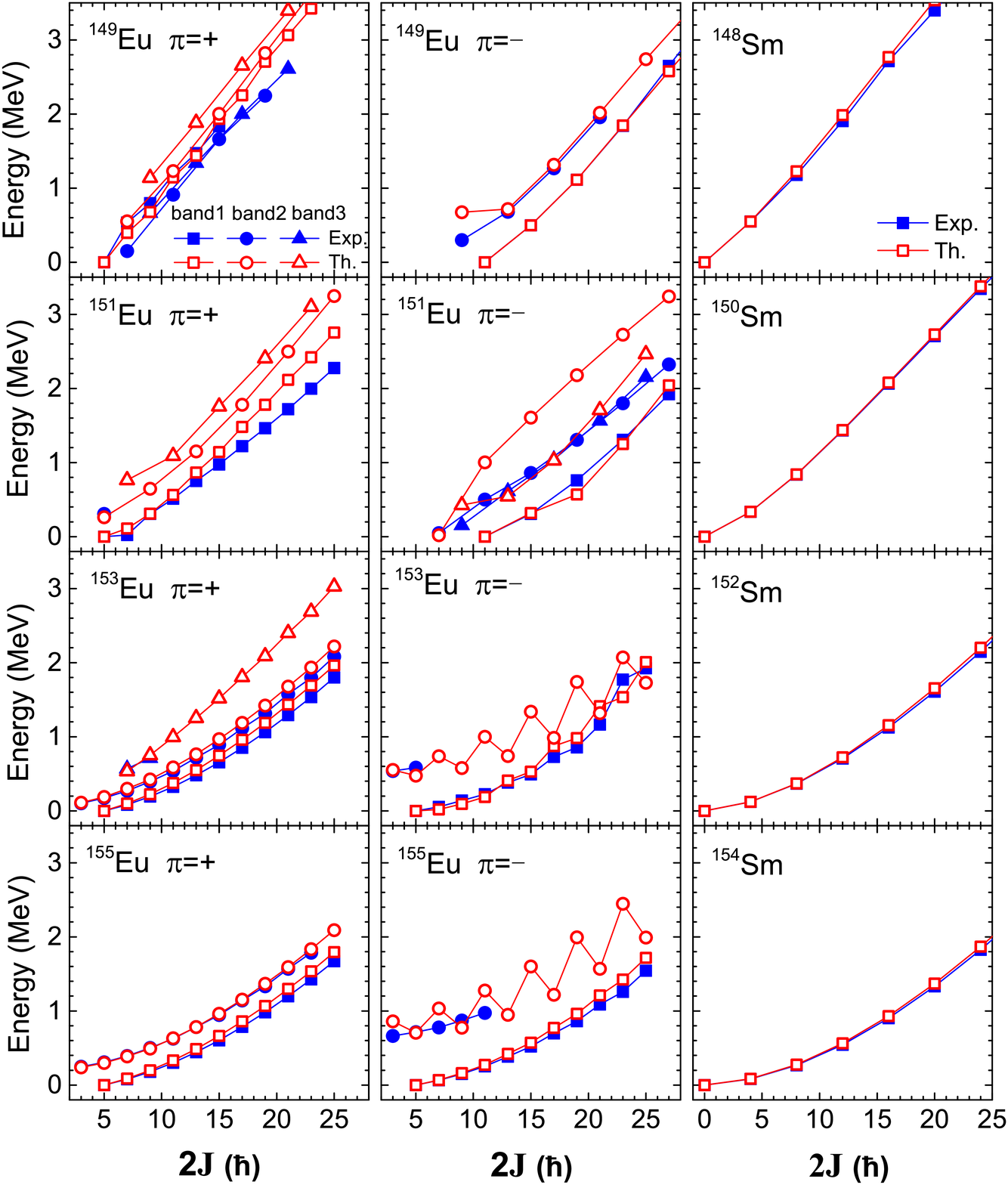}
\caption{\label{fig:spectrum}(Color online)  Low-energy positive-parity (left panels) and negative-parity (middle panels) bands of $^{149,151,153,155}$Eu isotopes as functions of  angular momentum, in comparison with the available data \cite{NNDC}. The excitation energies of the negative-parity states are shown relative to the corresponding lowest state. The ground-state bands of the adjacent even-even Sm isotopes are also shown in the panels on the right. The theoretical predictions for the odd-mass and even-even isotopes are obtained using the microscopic CQC Hamiltonian and 5DCH, respectively, based on the PC-PK1 energy density functional and a finite-range separable pairing force. The choice of the Fermi level and coupling strength ($\lambda$, $\chi$) in the CQC Hamiltonian: ($-4.80$, $8.80$), ($-8.60$, $13.4$), ($-8.05$, $11.0$), and ($-8.70$, $9.80$) reproduces the positive-parity bands of $^{149}$Eu, $^{151}$Eu, $^{153}$Eu, and $^{155}$Eu, respectively. The corresponding values of ($\lambda$, $\chi$) for the negative-parity bands of the four nuclei are ($-5.95$, $5.00$), ($-9.70$, $24.0$), ($-6.92$, $20.0$), and ($-8.00$, $19.6$), respectively. $\lambda$ is in units of MeV and $\chi$ in MeV/$b^2$.}
\end{figure}

Using Sm and Gd as the collective core nuclei, one can construct a microscopic core-quasiparticle coupling Hamiltonian for odd-mass Eu isotopes. The dynamics of the CQC Hamiltonian is determined by the energies, quadrupole matrix elements, and average pairing gaps corresponding to the spherical single-particle states of the unpaired nucleon, and collective excitation states of the two cores, which are calculated using the triaxial RHB method combined with the 5DCH. The Fermi level $\lambda$ and coupling strength $\chi$ of the core-quasiparticle quadrupole interaction are phenomenological parameters adjusted to reproduce the ground-state spin and/or the excitation energies of few lowest levels, separately for positive- and negative-parity states \cite{Quan17}.

Figure \ref{fig:spectrum} displays the low-energy positive- and negative-parity bands of $^{149,151,153,155}$Eu isotopes as functions of angular momentum, in comparison with available data \cite{NNDC}. The ground-state bands of the adjacent even-even Sm isotopes are also included. The calculated energy levels are grouped into bands according to the dominant $E2$ decay pattern. One notices that the theoretical results are in good agreement with experiment, not only for all the ground-state bands and lowest-lying negative-parity bands, but also for the one-quasiparticle excited bands. Only the positive-parity bands 2 and 3 in $^{149}$Eu, and the negative-parity band 2 in $^{151}$Eu are too high compared to the data, possibly because the model space does not include higher-order quasiparticle excitations. The calculated negative-parity band 2 of $^{153, 155}$Eu exhibits a staggering due to Coriolis coupling that is too strong, but this can be resolved by adding a magnetic dipole particle-core interaction term to the model Hamiltonian \cite{Prot97}. One also notices that the behavior of the excitation energies versus angular momentum for odd-mass Eu isotopes is consistent with that in the adjacent even-even Sm isotopes, namely from a nearly linear dependence characteristic for a spherical vibrator, to a parabolic dependence of an axial rotor as neutron number increases. We note that the Sm isotopes, and $^{152}$Sm in particular, were the first reported empirical example of a first-order QPT between a vibrator and axial rotor phases \cite{CZ.01}. A corresponding phase transition occurs in the odd-mass Eu isotopes. The negative-parity bands and the positive-parity excited bands exhibit a weak-coupling $\Delta J=2$ structure for $^{149, 151}$Eu, and rapidly change to the $\Delta J=1$ systematics of the strong-coupling limit for $^{153, 155}$Eu. 

\begin{figure}[htb!]
\centering
\includegraphics[width=8 cm]{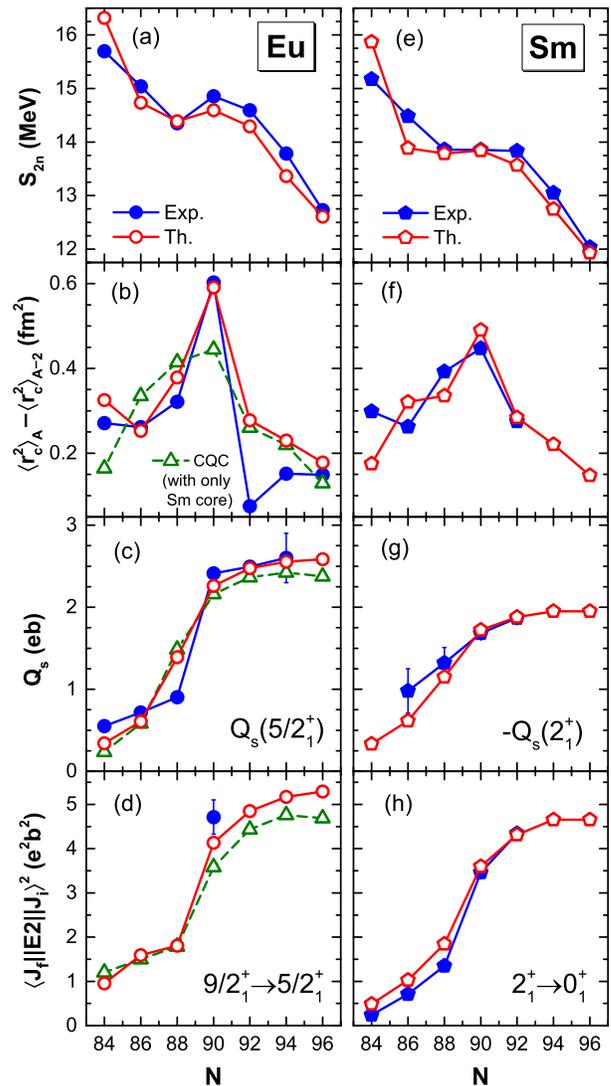}
\caption{\label{fig:obs}(Color online) Evolution of the two-neutron separation energies $S_{2n}$, isotope shifts of the ground-state charge radii $\langle r^2_c\rangle_{A}-\langle r^2_c\rangle_{A-2}$, spectroscopic quadrupole moments $Q_s$, and reduced transition matrix elements $\langle J_f||E2||J_i\rangle^2$, as functions of the neutron number in odd-mass Eu isotopes (a, b, c, d), and adjacent even-even Sm isotopes (e, f, g, h). The values calculated with the microscopic CQC Hamiltonian for Eu isotopes, and the 5DCH for Sm isotopes, are compared to available data \cite{NNDC}. Green triangles denote observables calculated with a CQC Hamiltonian that includes only Sm even-even cores.}
\end{figure}

To identify quantitative signatures of a possible shape QPT, we investigate the observables related to order parameters as functions of the control parameter -- nucleon number. A critical point of a QPT is characterized by a sudden change in the order parameter, even though one expects that in small systems with a finite number of particles the transition is, to a certain extent, smoothed out. In Fig.~\ref{fig:obs} we analyze the evolution with neutron number of the two-neutron separation energies $S_{2n}$, isotope shifts of the ground-state charge radii: $\langle r^2_c\rangle_{A}-\langle r^2_c\rangle_{A-2}$, spectroscopic quadrupole moments $Q_s$, and matrix elements $\langle J_f||E2||J_i\rangle^2$ for transitions to the ground state. The theoretical values are directly computed using the excitation energies and collective wave functions obtained with the CQC Hamiltonian. For comparison, we also include the isotopic dependence of the corresponding quantities in the adjacent even-even Sm nuclei, with the predictions of the 5DCH. Very similar values are also obtained for the even-A Gd isotopes. 

The agreement between the predictions and corresponding data is very good both for the even-even and odd-mass nuclei, especially considering that CDFT based on nuclear and pairing functionals are applicable over the entire chart of nuclides. In the context of the present study, an especially important result in Fig.~\ref{fig:obs} is that all considered observables present pronounced discontinuities at $N=90$. This points to the occurrence of a phase transition between spherical and quadrupole-deformed prolate shapes, and the $N=90$ isotones appear to be closest to the critical point. Furthermore, it is remarkable that the discontinuities of the order parameters for the odd-mass Eu isotopes are even steeper than those in the even-even Sm isotopes, particularly the isotope shifts and spectroscopic quadrupole moments. This means that the quadrupole interaction between the core and the unpaired fermion reinforces the QPT in 
odd-mass nuclei compared to the adjacent even-even isotopes. The enhancement of QPT in odd-mass systems was also discussed in Refs. \cite{FI11,Zhang13} by analyzing the contribution of deformation to two-neutron separation energies. Here we not only reproduce the sharper discontinuities in a microscopic calculation, but are also able to verify the enhancement of QPT in the odd-mass system by considering several observables.

\begin{figure}[htb!]
\centering
\includegraphics[width=8 cm]{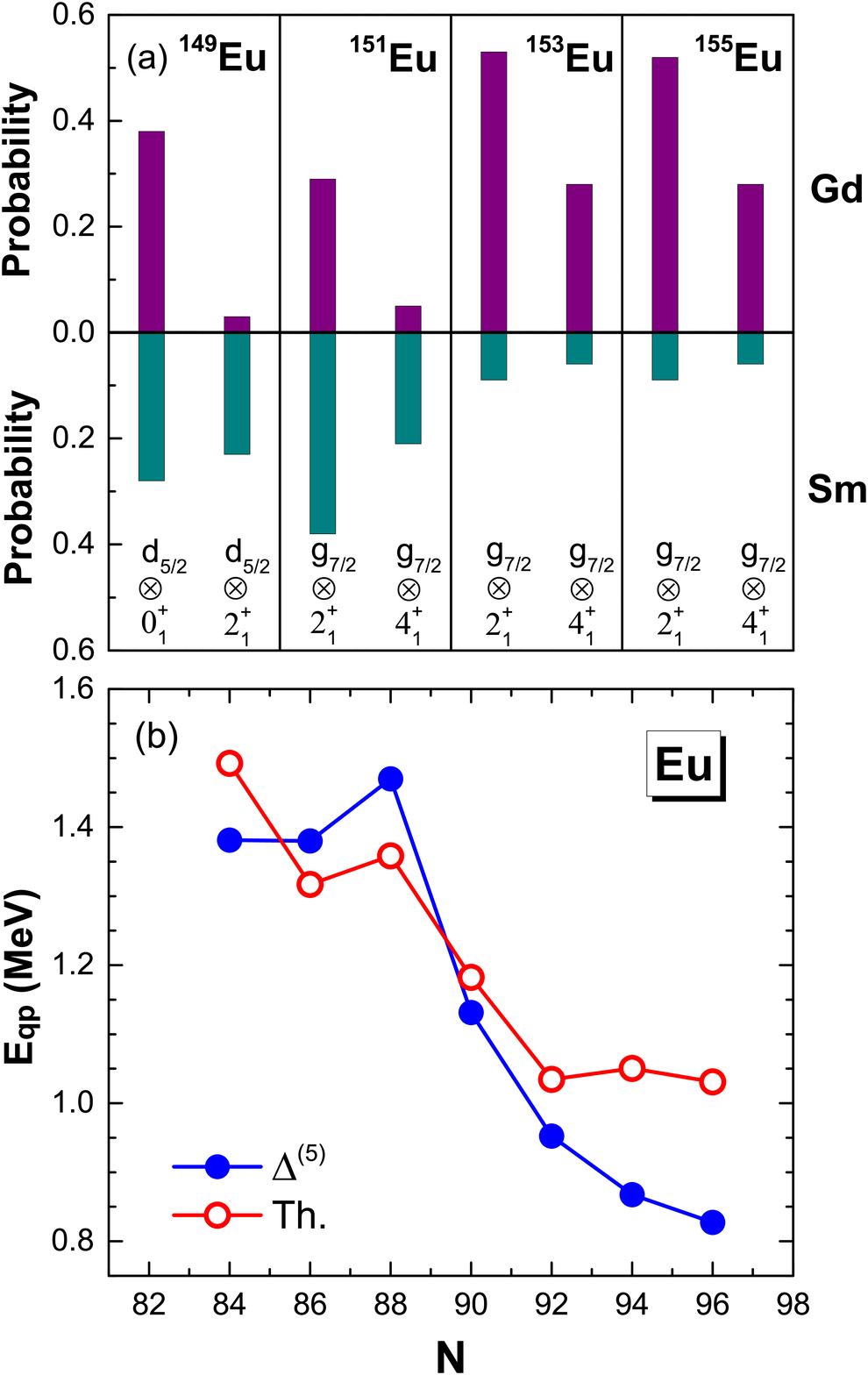}
\caption{\label{fig:wave} (Color online) Probabilities of the dominant configurations (a), and quasiparticle energies (b), for the ground states of Eu isotopes calculated with the CDFT-based CQC Hamiltonian. In panel (b) the experimental odd-even mass differences $\Delta^{(5)}$ are calculated using the five-point formula \cite{Bender00}.}
\end{figure}

The mechanism of the enhancement of QPT in odd-mass system is illustrated in Fig. \ref{fig:wave}, where we plot the probabilities of the dominant configurations in the ground states of the Eu isotopes and, in the lower panel, the corresponding quasiparticle energies. The quasiparticle energy of the ground state corresponds to the lowest eigenvalue of the CQC Hamiltonian, and it calculated as the difference between the total energy of the odd-mass nucleus and the average value of the energies of the two even-even cores. This is, of course, related to pairing correlations, and we compare the theoretical values to the empirical proton pairing gaps calculated using the five-point formula \cite{Bender00}. 
In Fig. \ref{fig:wave} (a) the ground state of $^{149}$Eu predominantly corresponds to the $2d_{5/2}$ spherical proton configuration, while those of $^{151, 153, 155}$Eu are dominated by the $1g_{7/2}$ configuration. One notices the rapid transition from configurations in which, because of shape fluctuations, the unpaired proton is almost equally coupled to both the Sm and Gd core low-spin yrast states, to ground states in the $N=90$ and $N=92$ Eu nuclei in which the proton is predominantly coupled to the Gd core. This is because with the increase of neutron number both Sm and Gd become markedly prolate deformed, but the Gd isotopes exhibit a slightly larger deformation, as evidenced by the spectroscopic quadrupole moments and matrix elements $\langle J_f||E2||J_i\rangle^2$ for transitions between yrast states. 
Consequently, the quadrupole core-proton interaction will favour coupling to the Gd core, and this
corresponds to a shape polarization effect that reinforces the QPT observed in the even-even isotopes. In addition, Gd exhibits a proton shell at $Z = 64$, and one expects a weaker proton pairing compared to Sm. The predominant coupling of the odd proton hole to the Gd core in Eu isotopes with $N \geq 90$ leads to the 
the sudden reduction of the ground-state quasiparticle energy at $N=90$ (cf. Fig.  \ref{fig:wave} (b)). This is also reflected in the pronounced kink observed for the two-neutron separation energies in odd-A Eu isotopes, as compared to the rather flat behavior of $S_{2n}$ around the critical point in the adjacent Sm and Gd isotopes (Fig. \ref{fig:obs} (a) and (e)). 

We note that a similar analysis of quantum shape phase transitions in odd-A Eu and Sm was performed in Ref.~\cite{Nomura16} using a framework based on EDFs and the particle-plus-boson-core coupling. The interacting boson model core Hamiltonian, as well as the single-particle
energies and occupation probabilities of the unpaired nucleon, are completely determined by constrained self-consistent
mean-field calculations for a specific choice of the EDF and paring interaction. The
strength parameters of the particle-core coupling are adjusted to reproduce selected spectroscopic
properties of the odd-mass system. Several quantities that can be related to quantum order
parameter were computed and their evolution
with neutron number analyzed. However, in contrast to the CQC Hamiltonian used in the present calculation, only the even-even Sm isotopes were considered as core nuclei, that is, the odd-fermion was only coupled to the corresponding $A-1$ core nucleus. With this choice of the boson core Hamiltonian one cannot analyze  the mechanism that, in the present study, enhances the first-order quantum phase transition in odd-mass systems. Namely, starting from the critical point at $N=90$ the odd-proton predominantly couples to the $A+1$ Gd core nuclei characterized by larger quadrupole deformation and weaker proton pairing correlations compared to the corresponding Sm isotopes. This effect is quantified in Fig.~\ref{fig:obs} where, with green triangles, we denote the isotope shifts of the ground-state charge radii, the spectroscopic quadrupole moments, and reduced transition matrix elements of odd-mass Eu isotopes calculated with a CQC Hamiltonian that is based only on Sm even-even cores. Obviously in this case the phase transition is less pronounced, and the agreement of the calculated ground-state quadrupole moments with data is not as good as in the case when the odd proton is allowed to couple to the $A+1$ Gd core.

%
%

In conclusion, a microscopic analysis of low-energy spectra and observables related to order parameters for a first-order nuclear QPT between spherical and axially deformed shapes in odd-mass Eu isotopes has been performed by solving a core-quasiparticle coupling Hamiltonian based on the PC-PK1 energy density functional. The calculated two-neutron separation energies, isotope shifts, spectroscopic quadrupole moments, and $E2$ reduced transition matrix elements are in very good agreement with available data, and exhibit sharper discontinuities at neutron number $N=90$ compared to those in adjacent even-even Sm and Gd isotopes.  The results indicate an enhancement of signatures of the first-order quantum phase transition in the odd-mass system. 
By analyzing the dominant configurations and quasiparticle energies of the ground state in Eu isotopes, the amplification of the QPT in the odd-mass system can be attributed to the shape polarization effect of the unpaired proton.

\bigskip
This work was supported in part by the NSFC under Grants No. 11475140,  No. 11575148, No. 11335002, and No. 11621131001, the Major State 973 Program of China No. 2013CB834400, the QuantiXLie Centre of Excellence, a project co-financed by the Croatian Government and European Union through the European Regional Development Fund - the Competitiveness and Cohesion Operational Programme (KK.01.1.1.01), and the Research Fund for the Doctoral Program of Higher Education under Grant No. 20110001110087.



\end{document}